\documentclass[twoside]{article}
\usepackage{arxiv}
\usepackage[utf8]{inputenc} 
\usepackage[T1]{fontenc}    
\usepackage{makecell}
\usepackage{booktabs}       
\usepackage{multirow}
\usepackage{siunitx}
\usepackage{hyperref}       
\usepackage{url}            
\usepackage{amsfonts}       
\usepackage{nicefrac}       
\usepackage{microtype}      
\usepackage{fancyhdr}       
\usepackage{graphicx}       

\title{The 2021 Tokyo Olympics Multilingual News Article Dataset
}

\author{
  Erik Novak, Erik Calcina, Dunja Mladenić, Marko Grobelnik \\
  Department for Artificial Intelligence \\
  Jožef Stefan Institute \\
  Ljubljana, Slovenia\\
  \texttt{\{name.surname\}@ijs.si} \\
}

\begin{document}

\flushbottom
\maketitle

\begin{abstract}

In this paper, we introduce a dataset of multilingual news articles covering the 2021 Tokyo Olympics. A total of 10,940 news articles were gathered from 1,918 different publishers, covering 1,350 sub-events of the 2021 Olympics, and published between July 1, 2021, and August 14, 2021. These articles are written in nine languages from different language families and in different scripts. To create the dataset, the raw news articles were first retrieved via a service that collects and analyzes news articles. Then, the articles were grouped using an online clustering algorithm, with each group containing articles reporting on the same sub-event. Finally, the groups were manually annotated and evaluated. The development of this dataset aims to provide a resource for evaluating the performance of multilingual news clustering algorithms, for which limited datasets are available. It can also be used to analyze the dynamics and events of the 2021 Tokyo Olympics from different perspectives. The dataset is available in CSV format and can be accessed from the CLARIN.SI repository.

\end{abstract}



\thispagestyle{empty}


\section*{Background \& Summary}

Through news articles, we can learn about global events. Different publishers report the same event from various perspectives, highlighting what they find essential for their audience. Depending on the publisher's country, articles may be in different languages and may reflect the writer's biases. Thus, news articles are key for identifying world events, their coverage, and their global significance. To analyze these aspects, we need effective methods to group multilingual news articles based on their events, which typically involve similar entities (who/what was involved), time (when it happened), and place (where it happened).

There is a scarcity of multilingual news datasets for developing methods and models to group articles reporting on the same event. Table~\ref{tab:existing-datasets} lists the datasets suitable for news-related tasks. Most existing news datasets support research in classifying news articles into topics or domains~\cite{Lang1995-bf,Lewis1997-zu,Gulli2004-ru,Greene2006-qm,Lewis2004-kt,Wayne2001-zc,Norregaard2019-zc,Pelicon2020-qh,Gruppi2020-ys,Gebhard2020-fe}. Datasets like 20 Newsgroups~\cite{Lang1995-bf}, AG News~\cite{Gulli2004-ru}, BBC News Archive~\cite{Greene2006-qm}, and POLUSA~\cite{Gebhard2020-fe} contain articles annotated with a few labels, usually corresponding to the news domain (e.g. Sports, Business, Tech, World). Others, such as Routers-21578~\cite{Lewis1997-zu}, TDT2~\cite{Wayne2001-zc}, and News Category Dataset~\cite{Misra2021-bv,Misra2022-uq}, have annotated articles suitable for topic detection. Recent datasets, NELA-GT-2018~\cite{Norregaard2019-zc} and NELA-GT-2019~\cite{Gruppi2020-ys}, have multiple label groups corresponding to specific news aspects, such as political bias, reliability, and transparency. While they help classify news into topics or domains, these datasets are not necessarily suitable for identifying news events, as the labels are often too broad and invariant to the temporal and geographical information of the article. 

Furthermore, most datasets contain primarily English articles, making them unsuitable for multilingual tasks. Those that are multilingual are typically designed for other news-related tasks, such as discrimination (SETimes.HBS~\cite{Tiedemann2012-ks}) and alignment (Hashemi et al. dataset~\cite{Baradaran-Hashemi2010-sb}) between languages, language simplification (SNIML~\cite{Hauser2022-ng}), and news summarization (MassiveSumm~\cite{Varab2021-jj}). Multilingual news datasets, such as CC News~\cite{Mackenzie2020-rt} and Babel Briefings~\cite{Leeb2024-ge}, are developed to analyze world events, cultural narratives, and more. In addition to datasets, news monitoring systems such as Event Registry~\cite{Leban2014-no} and the GDELT Project~\cite{Leetaru2013-rx} track news reported from various international sources and perform data analysis. Both systems provide detailed event records, including date, location, and involving actors. The GDELT Project provides event information with the source article URLs used to detect the event. Event Registry provides the event information with it's news article metadata for analysis purposes.

Our research identified only one multilingual news dataset with articles annotated based on the events they cover. This dataset was prepared by Miranda et al.~\cite{Miranda2018-ct} for evaluating news stream clustering algorithms. It was initially gathered from the Event Registry~\cite{Leban2014-no} to assess a cross-lingual news similarity and event tracking method~\cite{Rupnik2016-dd}. The data set contains 34k news articles in various languages published between 2013 and 2015. The articles are grouped into 1.5k events covering various topics.

\begin{table}[ht]
    \centering
    \caption{Existing news article datasets grouped based on their expected usage. A label group is a set of labels corresponding to a particular news aspect, such as political bias, reliability, or transparency. Both news event tracking systems have an increasing number of articles and clusters, denoted with the star ($\star$) symbol.}
    \label{tab:existing-datasets}
    \centerline{
        \resizebox{1.\textwidth}{!}{
            \begin{tabular}{lcccc}
                \toprule
                \thead[l]{\textbf{Corpus name}}                           & 
                \thead[c]{\textbf{News publication}\\\textbf{time range}} & 
                \thead[c]{\textbf{Language(s)}}                           & 
                \thead[c]{\textbf{No. articles}}                          & 
                \thead[c]{\textbf{No. clusters}\\\textbf{or labels}}      \\
                
                \midrule
                \multicolumn{5}{l}{\textsc{News event clustering datasets}} \\
                \midrule
        
                Miranda et al. Dataset~\cite{Miranda2018-ct}              & Dec 2013 - Avg 2015 & English    & 21k    & 808 \\
                                                                          &                     & Spanish    & 6.7k   & 554 \\
                                                                          &                     & German     & 6.1k   & 490 \\
                                                                          &                     & Chinese    & 450    & 9  \\
                                                                          &                     & Slovenian  & 37     & 3   \\
                                                                          &                     & Croatian   & 13     & 2   \\
                                                                          &                     & French     & 61     & 2   \\
                                                                          &                     & Russian    & 231    & 1   \\
                                                                          &                     & Italian    & 88     & 2   \\
                                                                          
                \midrule
                \multicolumn{5}{l}{\textsc{News classification datasets}} \\
                \midrule
        
                News Category Dataset~\cite{Misra2021-bv,Misra2022-uq} & Jan 2012 - Sept 2022   & English        & 210k    & 42 \\
                POLUSA dataset~\cite{Gebhard2020-fe}               & Jan 2017 - Aug 2019    & English        & 0.9M   & 18  \\
                NELA-GT-2019~\cite{Gruppi2020-ys}                  & Jan 2019 - Dec 2019    & English        & 1.12M  & 7 label groups* \\
                NELA-GT-2018~\cite{Norregaard2019-zc}              & Feb 2018 - Nov 2018    & English        & 713k   & 8 label groups* \\
                BBC News Archive~\cite{Greene2006-qm}              & 2004 - 2005            & English        & 2k     & 5   \\
                AG News~\cite{Gulli2004-ru}                        & 2004                   & English        & 1M     & 4   \\  
                TDT2 Multilanguage Text Corpus~\cite{Wayne2001-zc}  & Jan 1998 - June 1998   & English       & 53.6k  & 100 \\
                                                                   &                        & Chinese        & 18.8k  & 100 \\
                RCV1~\cite{Lewis2004-kt}                           & Aug 1996 - Aug 1997    & English        & 804k   & 53  \\
                20 Newsgroups~\cite{Lang1995-bf}                   & Apr 1993 - May 1993    & English        & 20k    & 20  \\
                Routers-21578~\cite{Lewis1997-zu}                  & Feb 1987 - Oct 1987    & English        & 21.5k  & 135 \\
                Pelicon et al. Dataset~\cite{Pelicon2020-qh}       & -                      & Croatian       & 2k     & 3   \\
                \midrule
                \multicolumn{5}{l}{\textsc{Other news datasets}} \\
                \midrule
                
                SNIML~\cite{Hauser2022-ng}                             & 2003 - 2022           & 6 languages    & 13.4k   & -     \\
                
                Babel Briefings~\cite{Leeb2024-ge}                     & Aug 2020 - Nov 2021   & 30 languages   & 4.7M    & -     \\
                CC News~\cite{Mackenzie2020-rt}                        & Jan 2017 - Dec 2019   & English        & 708k    & -     \\
                Hashemi et al. Dataset~\cite{Baradaran-Hashemi2010-sb} & Jan 2002 - Dec 2006   & English \& Persian       & 245k     & -     \\
                MassiveSumm~\cite{Varab2021-jj}                        & -                     & 92 languages   & 28.8M   & -     \\
                SETimes.HBS~\cite{Tiedemann2012-ks}                    & -                     & 3 languages    & 
                9k      & -     \\
        
                \midrule
                \multicolumn{5}{l}{\textsc{News event tracking systems}} \\
                \midrule
                
                Event Registry~\cite{Leban2014-no}                     & since 2014            & +60 languages  & $\star$ & $\star$  \\ 
                GDELT Project~\cite{Leetaru2013-rx}                    & since 1979            & +100 languages & $\star$ & $\star$ \\
                
                \bottomrule
                
            \end{tabular}
        }
    }
\end{table}

\begin{figure}[ht]
    \centering
    \includegraphics[width=\linewidth]{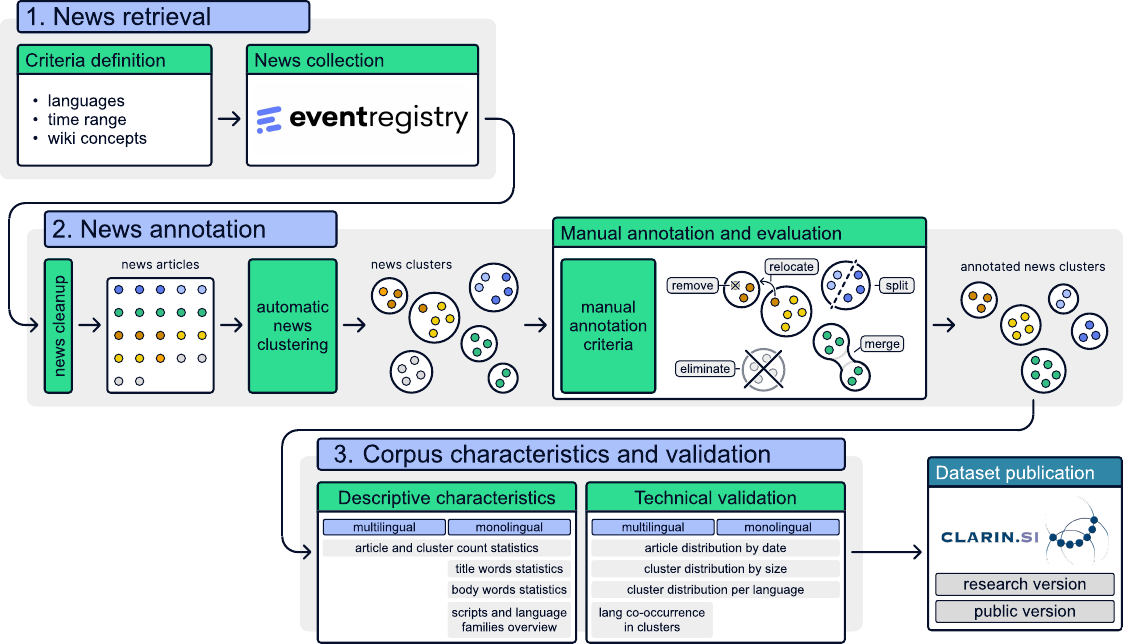}
    \caption[The schematic overview of the OG2021 development.]{The schematic overview of the OG2021 development.}
    \label{fig:og2021-schematic-overview}
\end{figure}

Because of the low number of events in the wide time range of the Miranda et al. dataset, it may not be suitable for developing approaches focused on high-frequency events, i.e., events that happen close in temporal proximity and possibly in similar locations. To address this gap, we created a novel multilingual news dataset named OG2021, containing articles reporting on the 2021 Tokyo Olympics. The articles are written in multiple languages and annotated based on the events they report; articles on the same event have the same annotation. The Olympics, spanning 18 days, presented a dense array of sub-events happening simultaneously, including articles with temporal, geographical, and contextual similarities that may challenge separation. Figure~\ref{fig:og2021-schematic-overview} illustrates the schematic overview of the novel OG2021 dataset preparation, including news article retrieval, annotation, and technical evaluation. The dataset was primarily developed to evaluate online multilingual news clustering algorithms in a high-frequency event setting. However, it can also be used to analyze the dynamics and events of the 2021 Tokyo Olympics, including cultural and linguistic differences in the articles and challenges faced by the organizers and competitors, such as COVID regulations.

\section*{Methods}

This section focuses on the approach used to create the OG2021 dataset. It first describes the news retrieval, followed by their annotation process.

\subsection*{News Retrieval}

Before we can begin any data processing, we must first retrieve the relevant news articles. This section describes the defined criteria and the approach taken to retrieve these articles.

\subsubsection*{Criteria definition}

To retrieve the appropriate articles, we first defined the criteria that the articles must follow. The requirements consist of three conditions: (1) the languages in which the article must be written, (2) the publication time, which must be within a predefined time range, and (3) the contextual concepts the article must include. Table~\ref{tab:criteria-conditions} shows the overview of the defined conditions. Below is a detailed description of each criterion.

\begin{table}[ht]
    \centering
    \caption[The news retrieval criteria conditions.]{The news retrieval criteria conditions.}
    \label{tab:criteria-conditions}
    \centerline{
        \resizebox{1.\textwidth}{!}{
            \begin{tabular}{cll}
                \toprule
                \thead[c]{\textbf{No.}}               & 
                \thead[l]{\textbf{Condition type}}    & 
                \thead[l]{\textbf{Condition values}}  \\
                
                \midrule
                
                1 & Languages              & \makecell[lt]{English, Portuguese, Spanish, French, Russian, German, Slovenian, Arabic, Chinese} \\
                2 & Publication time range & July 1, 2021 - August 14, 2021 \\
                3 & Contextual concepts    & \makecell[lt]{\texttt{Olympic Games}, \texttt{Japan} and at least one of \texttt{basketball}, \texttt{sports climbing},\\\texttt{swimming}, \texttt{judo}, \texttt{rowing}, \texttt{skateboarding}, or \texttt{table tennis}} \\
                
                \bottomrule
            \end{tabular}
        }
    }
\end{table}

\vspace{\baselineskip}
\noindent
\textit{Languages}. 
We aim to include articles in languages from diverse language families and scripts to represent a wide range of linguistic features. Following Miranda et al.'s dataset, we chose English, Spanish, German, French, Russian, Chinese, and Slovenian. We also include Portuguese and Arabic to increase linguistic diversity. The selected languages cover different language families (Germanic, Italic, Slavic, Semitic, and Sinitic) and are written in different scripts (Latin, Cyrillic, Arabic, and Chinese).

\vspace{\baselineskip}
\noindent
\textit{Publication time range}. 
To capture relevant articles, we limited our selection to those published between July 1, 2021, and August 14, 2021. This time range includes articles from three weeks before the start of the 2021 Tokyo Olympics (July 23, 2021) to one week after its conclusion (August 8, 2021). This allowed us to gather articles covering the events leading up to the opening as well as post-event coverage.

\vspace{\baselineskip}
\noindent
\textit{Contextual concepts}. 
The articles' contextual concepts must be related to the 2021 Tokyo Olympics. Because of this, we decided the articles must first be related to the general concepts \texttt{Olympic Games} and \texttt{Japan},  which will limit the scope of retrieved articles. To further narrow the focus, we concentrate on seven different sports: \texttt{basketball}, \texttt{sports climbing}, \texttt{swimming}, \texttt{judo}, \texttt{rowing}, \texttt{skateboarding}, and \texttt{table tennis}. The sports were chosen for their diversity in competition organization and duration, as well as their historic presence in previous Olympic Games. Notably, \texttt{sports climbing} and \texttt{skateboarding} debuted in the 2021 Olympic Games.

\subsubsection*{News collection}

With the defined article criteria, we can now retrieve the relevant news articles. To do this, we use Event Registry~\cite{Leban2014-no}, the system that collects news articles from thousands of publishers, clusters them into news events, and enriches news articles and event clusters by extracting the named entities mentioned. It also links the article's textual components to corresponding Wikipedia pages through a process called  wikification~\cite{Brank2017-fb,Brank2018-rx}. Due to the structure of Wikipedia, where different-language Wikipedia pages corresponding to the same concept are linked, the wikification process identifies the Wikipedia pages in both the article's language and in English, if available.

The Event Registry system has a dedicated API (\url{https://www.newsapi.ai/}), which enables the retrieval of news articles and event cluster metadata. The retrieval can be done through multiple endpoints, which accept various parameters, including the language of the article, the publication date range, and the Wikipedia concepts the articles must relate to. Although the system provides clustered news articles, we used the API to collect only the news articles, as the clusters created by the system do not have the structure we are targeting.

Using the defined criteria, we developed code to retrieve the relevant articles automatically via the Event Registry API. We split the time range into days, then retrieved the articles published on a given day that were written in one of the selected languages and contained both the \texttt{Olympic Games} and \texttt{Japan} Wikipedia concepts, as well as at least one of the sports concepts listed in the criteria definition. The API allows using English Wikipedia concepts to retrieve articles in other languages, eliminating the need for prior translation.

Through this process, we retrieved 36k articles that matched the defined criteria. In addition, some articles cover also concepts not defined in the criteria, such as additional sports and issues relating to COVID. Therefore, the dataset covers a broader range of concepts. The articles were then sent for annotation.

\subsection*{News Annotation}

Once the news articles were retrieved, we proceeded with the annotation process. The goal is to annotate articles based on the events they reported on, i.e., ensuring that articles reporting on the same event received the same annotation. We first describe how we prepare and clean the retrieved news articles. Afterward, we present the annotation process, which comprises automatic news clustering and the subsequent manual annotation and evaluation.

\subsubsection*{News cleanup}

The retrieved articles have various attributes, both original and enriched by the Event Registry. To provide only the original information, we include the following attributes: its unique ID, its title and body text, the publication datetime, its language, its published URL address and the name of the source publisher. We then split the articles into seven datasets, each corresponding to one of the sports concepts used during the retrieval process. Each dataset is then processed in the automatic clustering and manual annotation and evaluation, in that order.

\subsubsection*{Automatic news clustering}
We apply an online multilingual news clustering algorithm~\cite{Novak2021-oo} to automatically cluster each dataset corresponding to one of the sport's concepts. This algorithm, a single-pass clustering approach, processes the collected datasets by representing each article using its content embedding, the set of extracted named entities, and its publication datetime. The content embedding is created using the SBERT~\cite{Reimers2019-qr,Thakur2020-rf} language model, trained to generate contextual embeddings appropriate for pairwise sentence similarity, while the named entities were extracted using the WikiNEuRal~\cite{Tedeschi2021-go} multilingual named entity extraction model. 

The representations for event clusters are created as aggregates of content embeddings, entity sets, and publication datetime values of articles within the clusters. The algorithm measures the content similarity, ratio of overlapping entities, and temporal proximity between the article and all clusters, checking if all three values are above the set thresholds defined at the beginning of the clustering process. The article is placed into the best event cluster that meets these criteria. If no such cluster exists, a new one is created containing the current article. The algorithm's hyper-parameters (thresholds) were chosen to optimize precision while maintaining a reasonable recall score. This intentional selection results in more clusters containing only the most similar articles, potentially reducing the inclusion of articles that should not be clustered together.

The automatic news clustering resulted in around 16k event clusters containing an average of 2-3 articles. These clusters are then processed through manual annotation in the next step.

\subsubsection*{Manual annotation and evaluation}
Following the automatic news clustering process, the datasets underwent a manual evaluation and annotation process. Each dataset related to a sports concept was assessed using the following procedure.

\vspace{\baselineskip}
\noindent
\textit{Cluster representation}. To represent each automatically generated cluster, a table was created to display the included article values. Each table row contains the current cluster ID, the unique article ID, the article publication datetime, the article language, as well as its title and body. Notably, the article's URL and source were intentionally omitted from the table, as they do not contain relevant event information. 

\vspace{\baselineskip}
\noindent
\textit{Manual annotation criteria}. The annotation criteria is designed to categorize articles based on their responses to fundamental journalistic questions: who, what, where, when, and how. The first four questions yield objective answers, while the response to how can involve subjectivity. Because of this, alignment with the how question is considered an optional criterion. The annotators are tasks to group articles based on the responses to these questions. In addition, articles reflecting the same event from different perspectives (e.g. two articles describing the finals of a sport event, where one focuses on the gold and the other on the silver medalists) should be in the same event cluster, although they might not have the same responses to the above questions.

\vspace{\baselineskip}
\noindent
To further improve the quality of the dataset, annotators are provided with additional guidelines:

\begin{itemize}
    \item Articles should predominantly focus on a singular event, situation, or key information.
    \item Articles must avoid resembling “click-bait,” meaning they should not include phrases like “watch now” or “click here to see more.” If they do, the article should be excluded from the dataset.
    \item Articles should not primarily focus on presenting a schedule of events occurring on a specific day. If they do, the article should be excluded from the dataset.
\end{itemize}

\noindent
The intentionally ambiguous criterion aims to create diverse clusters of articles in various languages. Additionally, the criterion allows validation of an article based solely on its title, provided that the title is informative enough to provide the correct annotation.

\vspace{\baselineskip}
\noindent
\textit{Manual evaluation process}. Given the evaluation criteria, the annotators were tasked with reviewing the clusters and annotating the articles. If an annotator encountered a text that was not comprehensible, they were permitted to utilize machine translation services such as Google Translate (\url{https://translate.google.com/}). Annotators had the flexibility to remove specific articles from the dataset, relocate articles between clusters, merge, divide, and eliminate entire clusters. Each action taken was recorded and contributed to the creation of the final dataset.

After the manual annotation and evaluation process for all datasets, the annotated datasets underwent a joint evaluation, focusing on the final annotation process that involved joining clusters reporting on the same event and removing any inappropriate articles that might have been overlooked during the initial annotation pass. Due to the removal of irrelevant news articles and event clusters, corresponding to the schedules and ``click-bait'' like content, the number of articles was reduced by around 26k articles. This shows how much content of that nature is generated in the short time span. Furthermore, the number of clusters was also significantly reduced due to the removal and merging of event clusters. The final dataset was then formatted and prepared for analysis and publication.

\section*{Data Records}

The OG2021 dataset consists of 10,940 news articles reporting on the 2021 Tokyo Olympics. Table~\ref{tab:og2021-dataset-variables} shows the variables describing each news article. Each article is described with its unique ID, its title and truncated body, the publication datetime, its language, its published URL address, and the source publisher. Each news article is also assigned a cluster ID, indicating articles reporting on the same sub-event during the 2021 Olympics. The news articles are written in nine different languages (English, Portuguese, Spanish, French, Russian, German, Slovenian, Arabic and Chinese), with the article’s publication datetime spanning between July 1, 2021 and August 14, 2021. 

\begin{table}[ht]
    \centering
    \caption[The OG2021 variables used in the released version.]{The OG2021 variables used in the released version. Each variable is presented with its name, type, and description. Due to copyright and legal restrictions, the truncated body of the news article is available only in the research version of the dataset.}
    \label{tab:og2021-dataset-variables}
    \centerline{
        \resizebox{1.\textwidth}{!}{
            \begin{tabular}{clll}
                \toprule
                \thead[c]{\textbf{No.}}           & 
                \thead[l]{\textbf{Variable name}} & 
                \thead[l]{\textbf{Variable type}} & 
                \thead[l]{\textbf{Description}}   \\
                
                \midrule
                
                1 & ID     & number & The unique ID of the news article. \\
                2 & TITLE  & string & The title of the news article. \\
                3 & BODY*  & string & The truncated body of the news article. \\
                4 & LANG   & string & \makecell[lt]{The language in which the article is written. It can be one of nine values:\\\texttt{eng}, \texttt{por}, \texttt{spa}, \texttt{fra}, \texttt{rus}, \texttt{deu}, \texttt{slv}, \texttt{ara}, \texttt{zho}.} \\
                5 & SOURCE & string & The news publisher's name. \\
                6 & PUBLISHED\_AT & date & \makecell[lt]{The date and time the article was published.\\Format: \texttt{YYYY-mm-DD HH:MM:SS}.} \\
                7 & URL & string & The URL location of the news article. \\
                8 & CLUSTER\_ID & string & \makecell[lt]{The ID of the cluster the article is a member of.\\Format: \texttt{cls-xxxxxxxx}, where \texttt{x} can be a number of character.} \\ 
                \bottomrule
            \end{tabular}
        }
    }
    
\end{table}

Table~\ref{tab:og2021-dataset-statistics} shows the statistics of the dataset, including the average number of words in the article's title and body, the average cluster size, along with their standard deviation. The most present language in the dataset is English, comprised of 4k articles, followed by Portuguese and Spanish. The least articles present in the dataset is in Chinese with a total of nine articles. Other languages contain between 200 and 1,000 articles. In total, all articles are grouped into 1,350 event clusters.

\begin{table}[ht]
    \centering
    \caption{The OG2021 dataset statistics. It shows the number of articles, the average number of words in the title and body, the number of clusters, and the average cluster size. For Chinese, we report the average number of characters in the title and body.}
    \label{tab:og2021-dataset-statistics}
    \centerline{
        \resizebox{1.\textwidth}{!}{
            \begin{tabular}{lccccccc}
                \toprule
                \thead[l]{\textbf{Language}}                      & 
                \thead{\textbf{Script}}                           & 
                \thead{\textbf{Language family}}                  & 
                \thead{\textbf{No. articles}\\(percent of total)} & 
                \thead{\textbf{Words in title}\\(mean and std)}   & 
                \thead{\textbf{Words in body}\\(mean and std)}    & 
                \thead{\textbf{No. clusters}}                     & 
                \thead{\textbf{Cluster size}\\(mean and std)}     \\
                
                \midrule
                
                All          & -        & -        & 10,940        & -      & -             & 1,350 & 8 (20) \\
                        
                \midrule
                
                English      & Latin    & Germanic & 4,009 (37\%)  & 11 (3) & 1,231 (1,147) & 729   & 5 (11) \\
                Portuguese   & Latin    & Italic   & 2,410 (22\%)  & 13 (3) & 527   (374)   & 368   & 7 (12) \\
                Spanish      & Latin    & Italic   & 2,049 (19\%)  & 13 (4) & 562   (427)   & 381   & 5 (9)  \\
                French       & Latin    & Italic   & 845   (8\%)   & 13 (4) & 565   (464)   & 170   & 5 (8)  \\
                Russian      & Cyrillic & Slavic   & 553   (5\%)   & 10 (3) & 301   (358)   & 152   & 4 (6)  \\
                German       & Latin    & Germanic & 516   (4\%)   & 9  (3) & 833   (1,011) & 100   & 5 (6)  \\
                Slovenian    & Latin    & Slavic   & 331   (3\%)   & 9  (3) & 450   (370)   & 102   & 3 (3)  \\
                Arabic       & Arabic   & Semitic  & 218   (2\%)   & 10 (3) & 405   (269)   & 71    & 3 (3)  \\
                Chinese$^{*}$& Chinese  & Sinitic  & 9     (0\%)   & 28 (7) & 3,402 (1,550) & 5     & 2 (1)  \\
                
                \bottomrule
            \end{tabular}
        }
    }
\end{table}

The OG2021 dataset is stored in a single CSV file, where each line corresponds to a single news article, and published on the CLARIN.SI repository. Furthermore, it is published in two versions: 

\begin{itemize}
    \item The public version~\cite{Novak2024-iq}. Due to legal restrictions, the public dataset does not contain the body of the articles. However, other article metadata is available, including its title and published URL address, that can be used to fetch the article's content. The dataset is publicly available and licensed under CC BY-NC-ND 4.0. 
    \item The research version~\cite{Novak2024-hs}. The research dataset contains all of the article attributes. The dataset is available for academic use and licensed under CLARIN.SI license ACA ID-BY-NC-INF-NORED 1.0, which requires the user to log into the CLARIN.SI repository via their academic institution.
\end{itemize}

\section*{Technical Validation}

In addition to the manual evaluation during the news annotation process, we also performed the technical validation to ensure the dataset indeed contains multilingual news articles within a high-frequency event setting.

\subsection*{Article distribution over time}
\begin{figure}[ht]
    \centering
    \includegraphics[width=\linewidth]{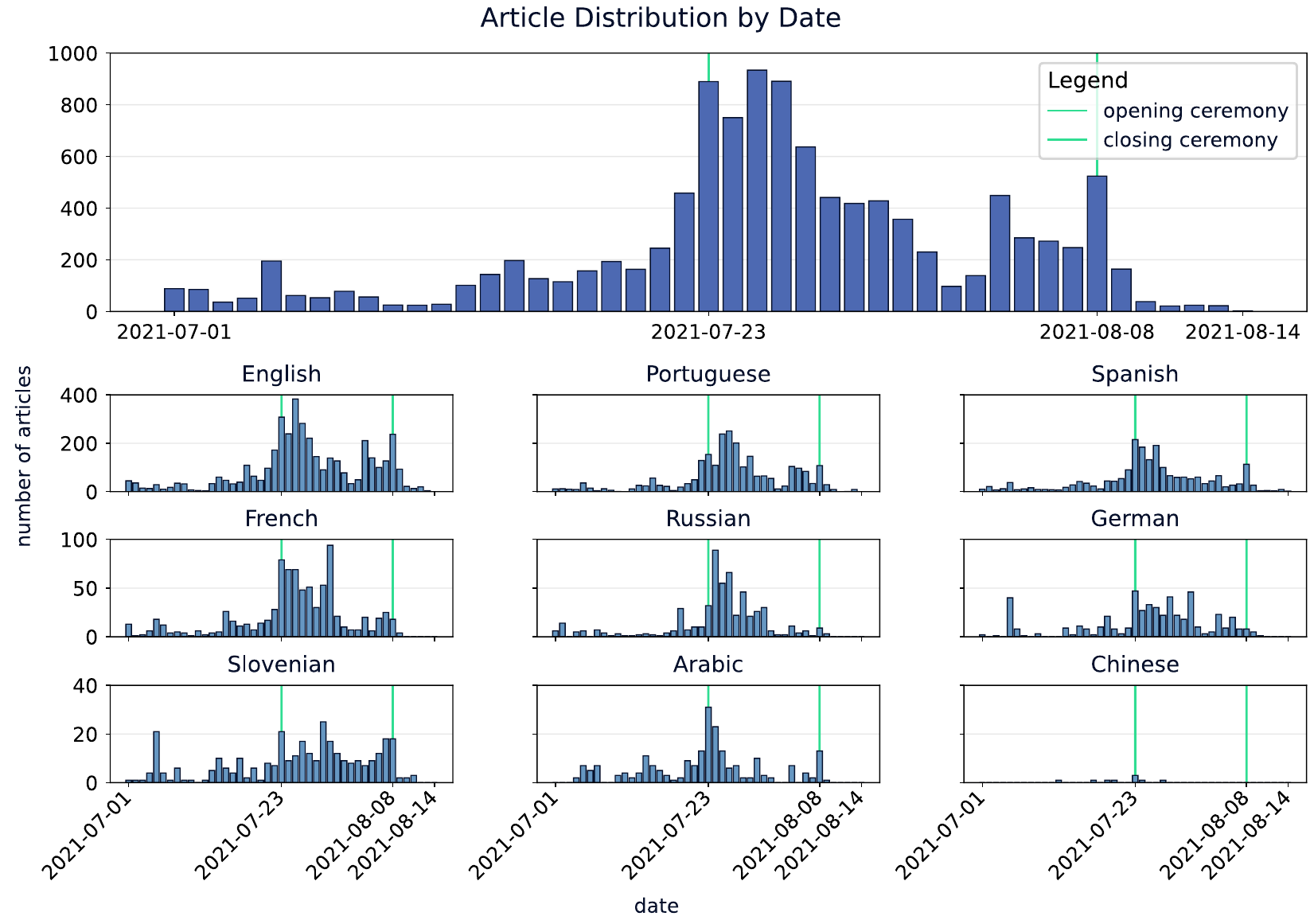}
    \caption[The OG2021 article distribution by date.]{The OG2021 article distribution by date. The majority of articles were published between the official start of the Olympic Games (July 23, 2021) and the official end of the Olympic Games (August 8, 2021).}
    \label{fig:og2021-article-distribution-by-date}
\end{figure}

Figure~\ref{fig:og2021-article-distribution-by-date} illustrates the article's distribution based on their publication datetime. The first articles were published on July 1, 2021. The peak concentration of articles occurs between July 23, 2021, and August 8, 2021, corresponding to the start and the end of the 2021 Tokyo Olympics event, respectively. The latest articles in the dataset were published on August 14, 2021. Furthermore, the article distributions for each language separately show a similar distribution, with the peak concentration of articles happening during the 2021 Olympics. 

\subsection*{Cluster distribution based on their size}
\begin{figure}[ht]
    \centering
    \includegraphics[width=\linewidth]{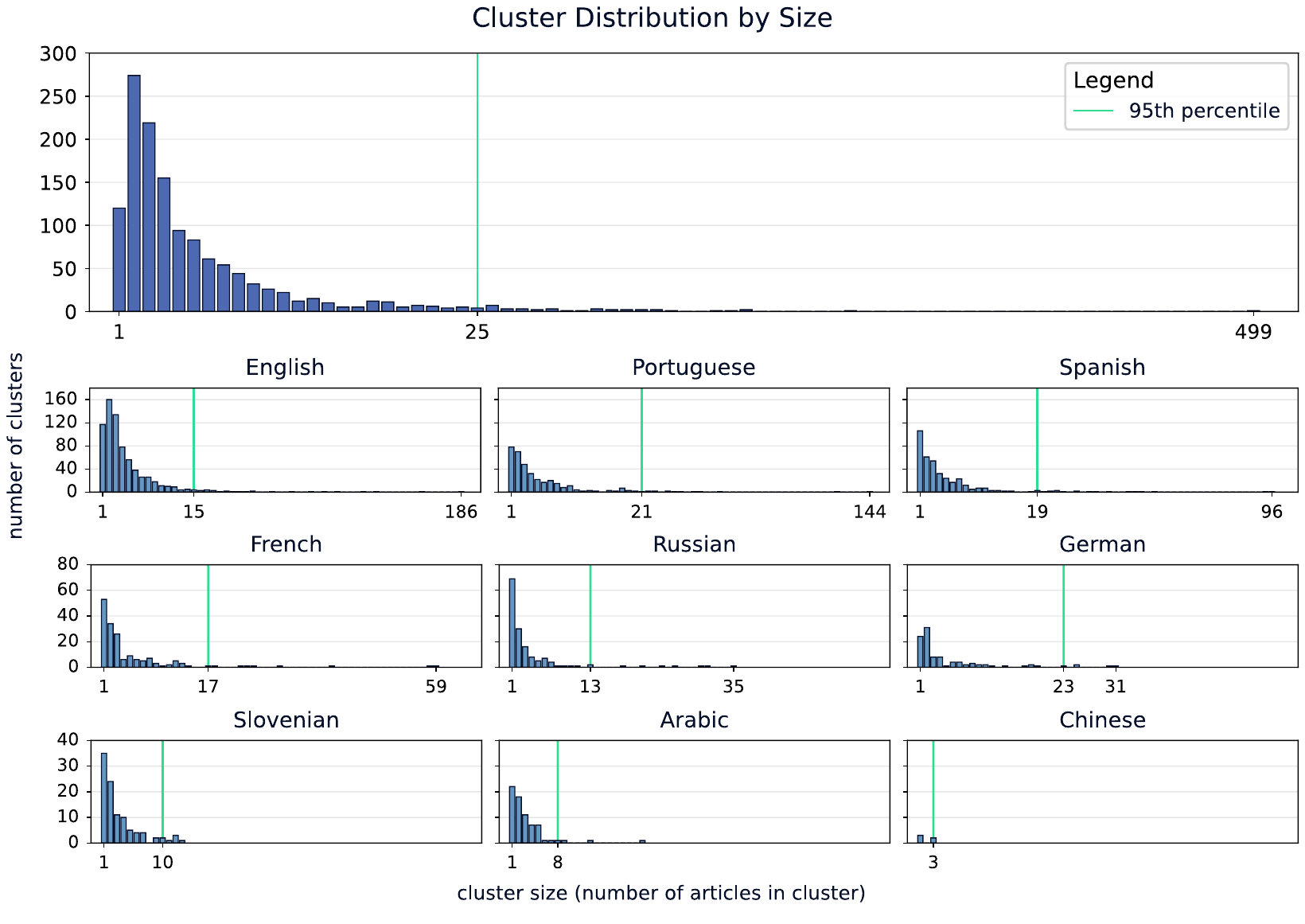}
    \caption[The OG2021 article distribution by size.]{The OG2021 article distribution by size. Globally, about 95\% of clusters contain 25 or fewer articles.}
    \label{fig:og2021-cluster-distribution-by-size}
\end{figure}

As mentioned before, the dataset is comprised of 1,350 distinct clusters, with an average size of eight articles per cluster. The distribution of clusters based on their size is shown in Figure~\ref{fig:og2021-cluster-distribution-by-size}. Looking at the whole dataset, around 95\% of clusters contain 25 articles or fewer. The highest number of clusters consists of only two articles, while the largest cluster, which corresponds to the 2021 Tokyo Olympics opening ceremony, contains 499 articles. Furthermore, isolating each language separately, the distribution varies across languages, showing that except for English and German, the highest number of clusters in each language contains a single article.

\subsection*{Cluster distribution per language}
\begin{figure}[ht]
    \centering
    \includegraphics[width=\linewidth]{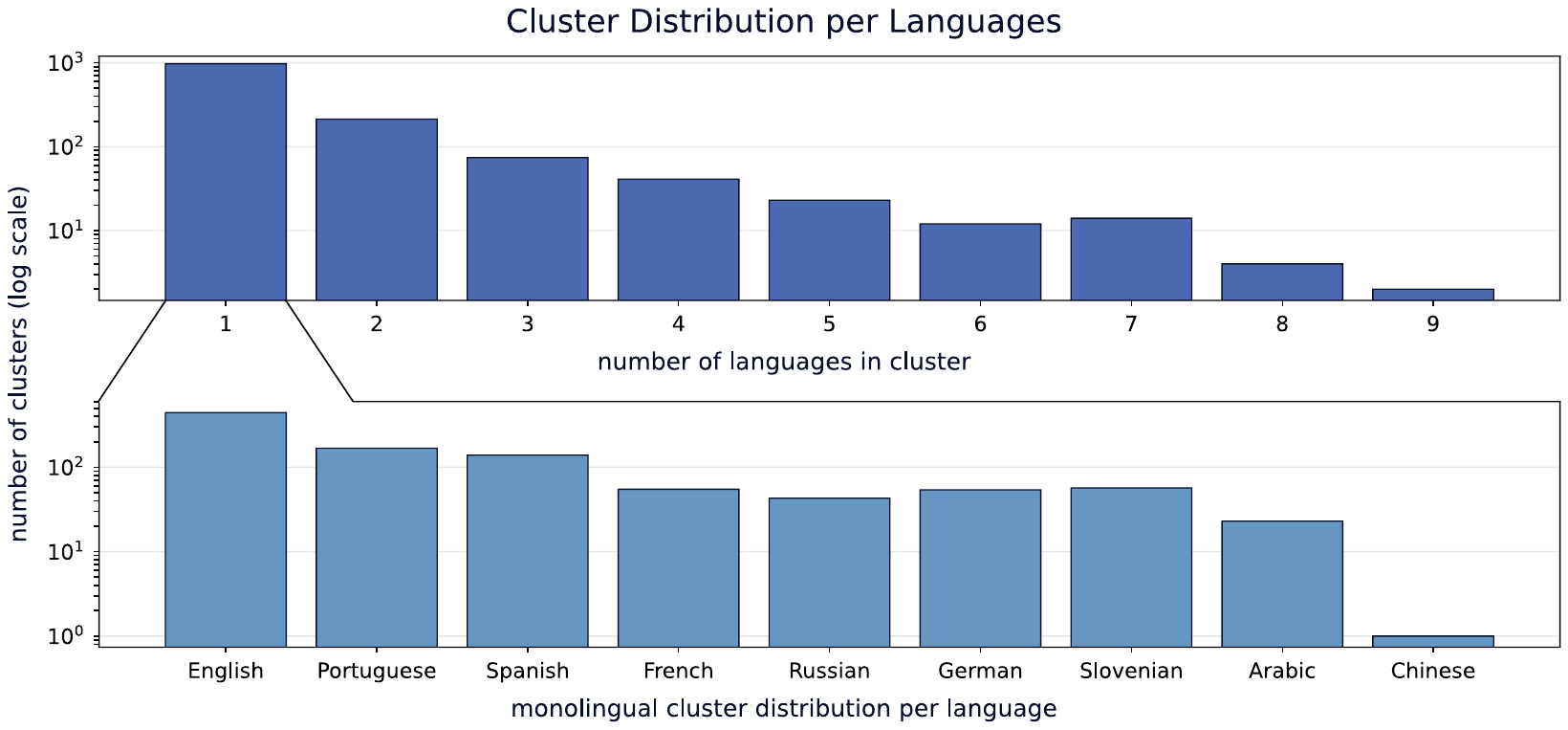}
    \caption[The OG2021 cluster distribution per language.]{The OG2021 cluster distribution per language. Almost 28\% of the clusters contain two or more languages. The lower graph shows the distribution of monolingual clusters across languages.}
    \label{fig:og2021-cluster-distribution-by-lang}
\end{figure}

Next, we counted the number of languages that appear in each cluster.Figure~\ref{fig:og2021-cluster-distribution-by-lang} shows the distribution of clusters based on the number of present languages. The majority of clusters, approximately 72\%, consist of articles written in a single language. The figure also shows the distribution of monolingual clusters, where the overall distribution per language roughly corresponds to the language's presence in the dataset. English has the highest number of monolingual clusters, followed by Portuguese and Spanish. French, Russian, German, Slovenian, and Arabic have approximately the same number of monolingual clusters. Chinese has no monolingual clusters, so its articles are always present with those from other languages.
Nonetheless, approximately 28\% of clusters include articles written in two or more languages. The highest number of languages found in a single cluster is nine, containing all the languages present in the dataset.

\subsection*{Language co-occurrence in clusters}
\begin{figure}[!ht]
    \centering
    \includegraphics[width=0.74\linewidth]{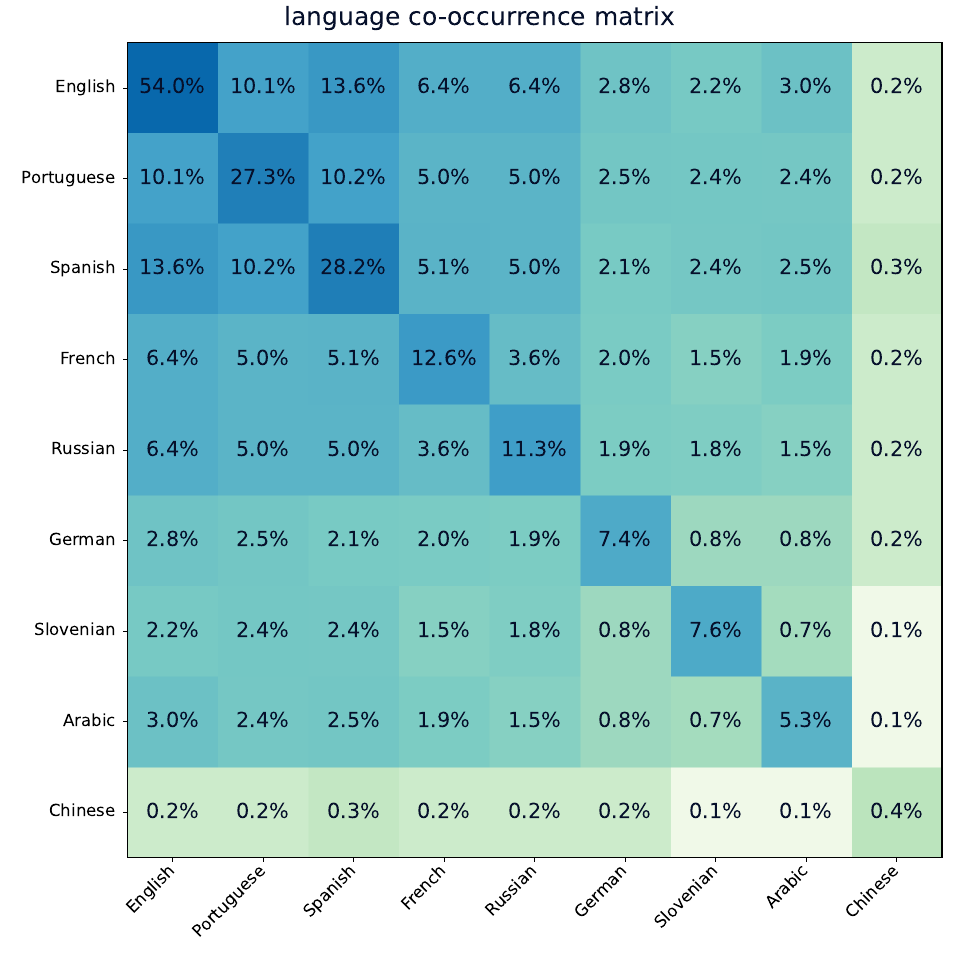}
    \caption[The OG2021 language co-occurrence in clusters.]{The OG2021 language co-occurrence in clusters. All language pairs appear together in at least one cluster.}
    \label{fig:og2021-language-cooccurrence}
\end{figure}

To further highlight the multilingual nature of the dataset, we computed the language co-occurrence across all clusters, as depicted in Figure~\ref{fig:og2021-language-cooccurrence}. The diagonal of the co-occurrence graph showcases the percentage of clusters in which the language is present. According to the calculations, any two languages co-appear in at least one cluster, equivalent to 0.1\% of the total number of clusters. The English, Portuguese, and Spanish language pairs co-occur in about 10\% of the clusters, while the remaining pairs exhibit co-occurrence ranging between 1-6\%. The exception is Chinese, which has a low presence in the dataset and consists of only nine articles.

\subsection*{Limitations and future work}

The raw news articles were collected using a list of concepts related to the 2021 Olympics and specific sports. Because of this, the dataset includes news articles that focus on the chosen sports rather than all sports at the event. For instance, \texttt{athletics} and \texttt{gymnastics} are less covered due to the defined retrieval conditions. Furthermore, the removal of ``click-bait'' and scheduling news articles significantly reduced the dataset size. These articles can introduce noise, which can be useful for developing methods to effectively group news articles or identify such types.

Considering these potential extensions, we plan to develop similar news article datasets focusing on future Olympic Games, starting with the 2024 Olympics in Paris. We would expand the retrieval conditions to include more languages and a broader list of sports. Additionally, we would include ``click-bait'' and scheduling articles, annotating them to reflect their content type.

\section*{Usage Notes}

To use the OG2021 dataset, the user must first download it from the CLARIN.SI repository~\cite{Novak2024-iq,Novak2024-hs}. Since the dataset is in CSV format, it can be opened using various programs and programming libraries.

\vspace{\baselineskip}
\noindent
The OG2021 dataset can be used to evaluate (online) multilingual news clustering algorithms, which is the main reason for creating it. The algorithm would process and group the news articles into event clusters. The created event clusters would then be compared with the dataset's cluster IDs to measure the algorithm's performance, which might include the standard and BCubed~\cite{Amigo2009-gd} variants of the F1, precision, and recall scores, cluster purity, and normalized mutual information. The performance metrics would then show how the algorithm fares in the high-frequency event setting presented within the dataset.

\vspace{\baselineskip}
\noindent
The dataset can also be used to analyze the dynamics and events of the 2021 Tokyo Olympics, including cultural differences and perspectives in reporting based on news publishers and the language used. Furthermore, it allows for analysis of the 2021 Olympics timeline, viewing the challenges faced by organizers and competitors and the solutions introduced.

\section*{Code availability}

The code version used to generate the OG2021 dataset is available on both Zenodo~\cite{Novak2024-wl}. 

\section*{Acknowledgements}

This work was supported by the Slovenian Research Agency and the European Union's Horizon 2020 project Humane AI Net [Grant No. 952026]. Furthermore, we would like to thank Anton Križnar and Matevž Matjašec for their contributions regarding the preliminary analysis of news articles, based on which we defined and executed the creation of the OG2021 dataset.

\bibliographystyle{unsrt}  
\bibliography{paperpile}

\end{document}